\documentclass[review]{elsarticle}
\usepackage{hyperref}
\usepackage[margin=1.1in]{geometry}
\usepackage[english]{babel}
\usepackage{float} 
\usepackage{natbib}
\usepackage{xcolor}
\usepackage{graphicx}
\graphicspath{ {./images/} }
\hypersetup{pdfauthor={Name}}

\newcommand{\rmsdBU}{RMSD\textsubscript{BU}}
\newcommand{\Angstrom}{\r{A}}

\makeatletter
\def\ps@pprintTitle{%
 \let\@oddhead\@empty
 \let\@evenhead\@empty
 \def\@oddfoot{\centerline{\thepage}}%
 \let\@evenfoot\@oddfoot}
\makeatother

\begin{document}

\begin{frontmatter}

\title{Advances to tackle backbone flexibility in protein docking}

\author[1]{Ameya Harmalkar}
\ead{ameya@jhu.edu}
\author[1,2]{Jeffrey J. Gray\corref{cor1}}
\ead{jgray@jhu.edu}

\cortext[cor1]{Corresponding author}

\address[1]{Department of Chemical and Biomolecular Engineering, Johns Hopkins University, Baltimore, MD, USA}
\address[2]{Program in Molecular Biophysics, Institute for Nanobiotechnology, and Center for Computational Biology, Johns Hopkins University, Baltimore, MD, USA}

\begin{abstract}

Computational docking methods can provide structural models of protein-protein complexes, but protein backbone flexibility upon association often thwarts accurate predictions. 
In recent blind challenges, medium or high accuracy models were submitted in less than 20\% of the ``difficult'' targets (with significant backbone change or uncertainty). 
Here, we describe recent developments in protein-protein docking and highlight advances that tackle  backbone flexibility. 
In molecular dynamics and Monte Carlo approaches, enhanced sampling techniques have reduced time-scale limitations. 
Internal coordinate formulations can now capture realistic motions of monomers and complexes using harmonic dynamics. And machine learning approaches adaptively guide docking trajectories or generate novel binding site predictions from deep neural networks trained on protein interfaces. These tools poise the field to break through the longstanding challenge of correctly predicting complex structures with significant conformational change.

\end{abstract}

\begin{keyword}
backbone flexibility $|$ protein-protein interactions $|$ conformational space 
\end{keyword}

\end{frontmatter}

\section*{Introduction}

Protein-protein interactions are involved in nearly all of the biological processes in human health and disease. Understanding the dynamics of binding and the structure of protein complexes at the molecular level can be instrumental in delineating biological mechanisms and developing intervention strategies. 
Computational protein-protein docking provides a route to predict the three-dimensional structures of protein assemblies or complexes from known structures of individual monomeric proteins \citep{Ritchie2008}.

\begin{figure}[tb]
\renewcommand{\familydefault}{\sfdefault}\normalfont
\centering
\includegraphics[width=14cm]{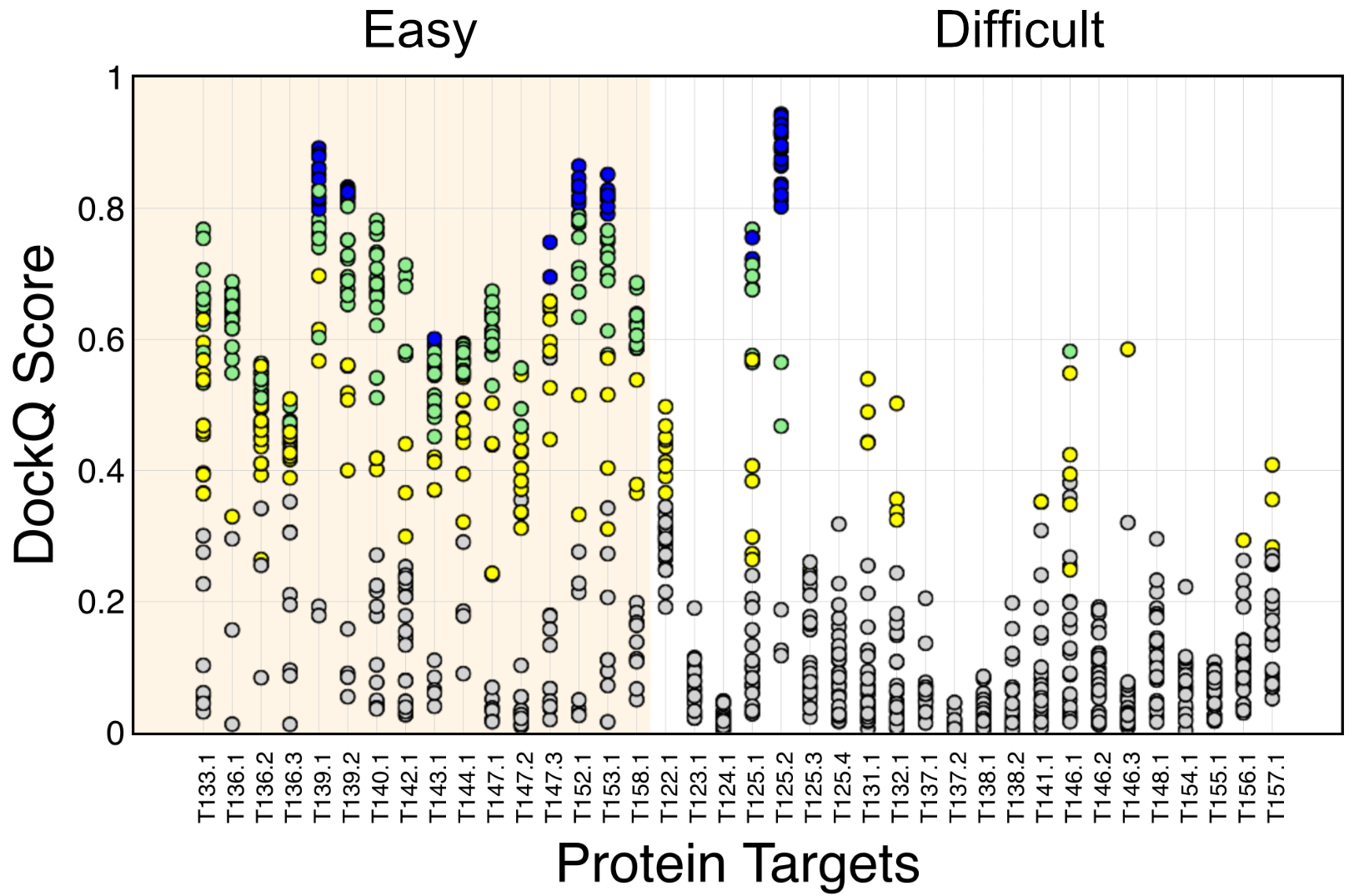}
\caption{\textbf{Performance of protein docking approaches on blind targets in CAPRI Rounds 38-46.} \citep{Lensink2019a,Lensink2019} Distribution of DockQ scores for the best model submitted by each predictor group (points) for each individual target (x-axis). DockQ measures a combination of intermolecular residue-residue contacts, interface RMSD, and ligand RMSD on a scale of 0 (incorrect) to 1 (matching the experimental structure) \citep{Lensink2019}. Targets are labelled by their CAPRI target number and, when needed, interface number (after the decimal). The targets are classified into rigid (easy) targets (high-homology monomer templates and under 1.2 \r{A} unbound-bound backbone motion, and flexible targets (poor template availability and/or over 1.2 \r{A} \rmsdBU). DockQ scores are color-coded by CAPRI model quality ranking: blue, high; green, medium; yellow, acceptable; gray, incorrect. Data graciously provided by Marc Lensink \citep{Lensink2019a, Lensink2019}. }
\label{fig:CAPRI}
\end{figure}

Docking methods are tested in the blind prediction challenge known as the Critical Assessment of PRediction of Interactions (CAPRI) \citep{Janin2003}, which in recent rounds pushed the field by including a wide array of target types such as transport proteins, higher order assemblies and host-virus interactions \citep{Lensink2018,Lensink2019a}. Out of the 28 protein-protein targets evaluated in CAPRI  over the past four years \citep{Lensink2019a, Lensink2019}, predictors achieved high quality structures for 11 ``easy'' targets, defined as those with little backbone motion (unbound to bound C$_{\alpha}$ root mean square deviation (\rmsdBU) of less than 1.2 \r{A} \citep{Vreven2015, Kundrotas2018}; Figure \ref{fig:CAPRI}). The remaining 17 targets were categorized as ``difficult'' (\rmsdBU\ over 2.2 \r{A} and/or poor monomer template availability). For these targets, predictors only achieved acceptable quality in 8 of 17 targets (47\%) and high quality in only 2 (12\%) \citep{Lensink2019a, Lensink2019}. Thus, the intrinsic flexibility of biomolecules still confounds the protein docking community at large.

In this review, we focus on the central docking challenge of capturing larger binding-induced conformational changes. 
We summarize progress by recent algorithms and frameworks, additionally augmented by growth in databases and computational power (CPU- and GPU-based). These new methods have achieved greater accuracy on more challenging targets and additionally yielded insight into binding mechanisms. We first present progress in binding site identification and then docking methods including molecular dynamics (MD) and Monte Carlo (MC) approaches, normal modes, and machine learning. Together, these techniques have helped better explore broader regions of conformational space and more thoroughly evaluate the energy landscape to improve protein-protein docking.

\begin{figure}[tb]
\renewcommand{\familydefault}{\sfdefault}\normalfont
\centering
\includegraphics[width=15cm]{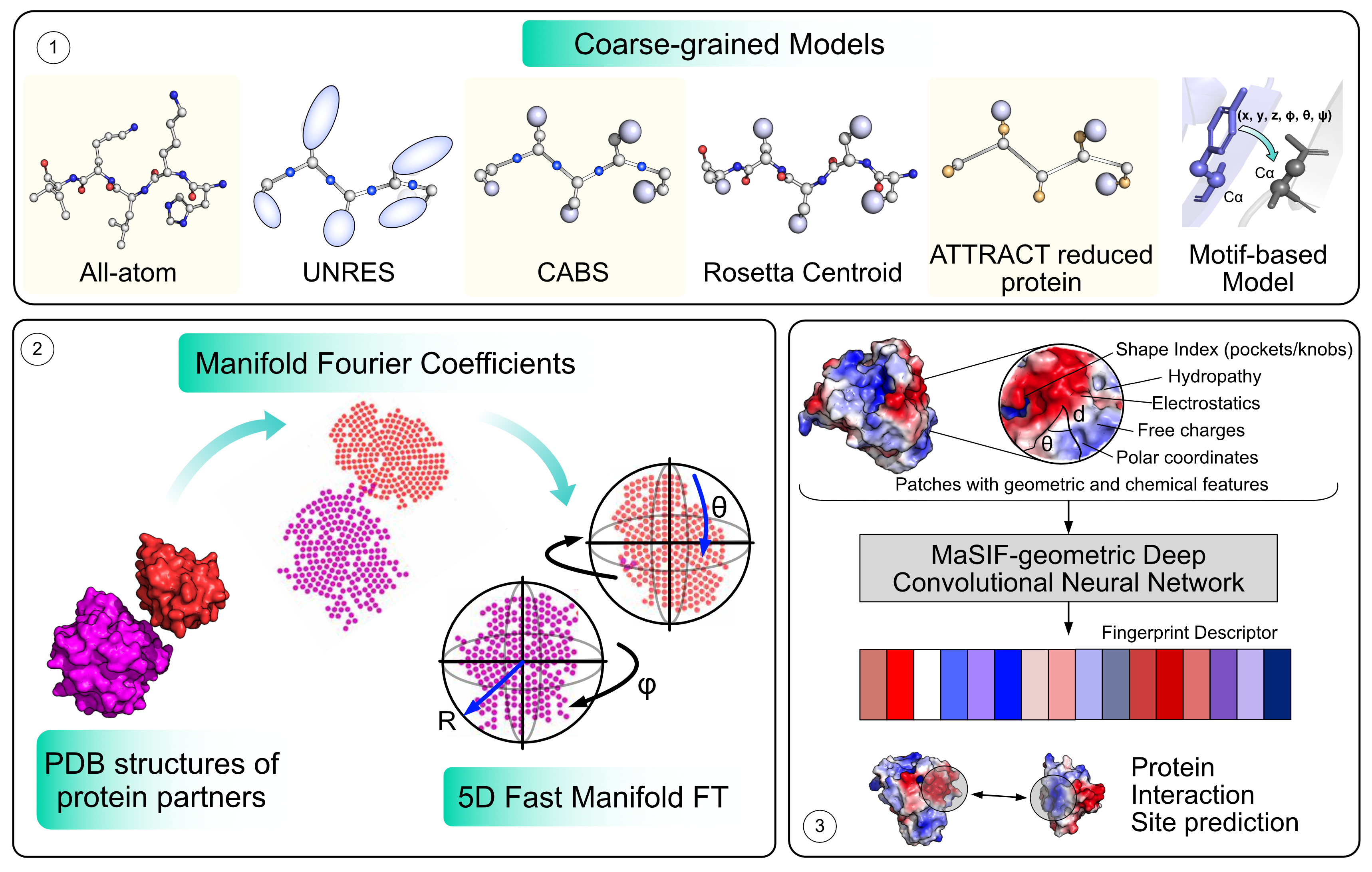}
\caption{\textbf{Reducing the degrees of freedom in protein docking.} 1.\textbf{Coarse-grained models} From left to right: Some approaches use all-atom representations (except solvent). The UNRES (united residue) model \citep{Liwo2014} represents the side chains as variable size ellipsoids attached to the C$_{\alpha}$ atom by peptide linkages and backbone N, C and O atoms are accounted with peptide-bond centers. CABS (C$_{\alpha}$,C$_{\beta}$ and sidechains) model adds a C$_{\beta}$ atom and approximates rest of the side chain by a single sphere. The Rosetta centroid model \citep{Lyskov2008} uses a CEN atom to represent the side chain while the backbone stays intact. The ATTRACT reduced protein model comprises of 2-3 atoms per residue with only C$_{\alpha}$ in the backbone and 1-2 atoms in the side chain \citep{Zacharias2005}. Knowledge-based model derived from residue pair transforms of protein motifs from bound complexes in the PDB \citep{Fallas2017,Marze2018}**. 2.\textbf{Fast manifold Fourier transforms (FMFT)}: The 5D FMFT method implicitly matches protein shapes over three translations and two rotations in Fourier space (adapted from Padhorny \textit{et al.} 2016 \citep{Padhorny2016}*). 3. \textbf{MaSIF} identifies binding sites using interface ``fingerprints'' in a geometric deep learning model \citep{Gainza2020}**. }
\label{fig:cg_fft_ml}
\end{figure}

\section*{Identifying putative binding sites: a global search}

To reduce the complexity of the immense conformation space of flexible proteins, coarse-grained models are frequently used to reduce the degrees of freedom (Figure \ref{fig:cg_fft_ml}). In the extreme, global docking approaches typically first treat protein partners as rigid bodies by restricting to six degrees of freedom (three rotational and three translational). A prime method to exhaustively sample the global 6D space is enumerating and scoring different rigid-body orientations on a dense grid. Approaches such as ClusPro \citep{Comeau2004, Kozakov2017}, ZDOCK \citep{Chen2003, Pierce2014}, PIPER \citep{Kozakov2006} and HexServer \citep{Macindoe2010} rely on the fast Fourier transform (FFT) correlation, which projects protein binding partners on a discretized three-dimensional grid. Conventional FFT approaches accelerate sampling only in the translational space and require new FFTs for every rotation. In 2015, Kazennov \textit{et al.} developed fast manifold Fourier transforms (FMFT) to search arrangements of two rigid bodies in a 5D manifold (Figure \ref{fig:cg_fft_ml}) \citep{Kazennov2015}. Relative to traditional FFT-based docking, FMFT accelerates calculations ten-fold \citep{Padhorny2016}*. 
Another shape-based approach is geometric hashing, which indexes point sets or curves to match geometric features under arbitrary transformations like translations, rotations or even scaling \citep{Smith2002}. Local 3D Zernike descriptor-based docking (LZerD), one of the top methods in CAPRI, projects 3D surfaces onto spheres to efficiently capture complementarity of protein surfaces
\citep{Venkatraman2009}. Some rigid-body approaches exploit data from chemical cross-linking experiments \citep{Vreven2018} or small-angle X-ray scattering (SAXS) \citep{Ignatov2018} to further improve discrimination of generated structures. These approaches provide fast, global exploration of the energy landscape, and in recent CAPRI rounds \citep{Lensink2019a, Lensink2019}, many predictors incorporated these approaches as the first step to identify putative binding patches, and they supplement with other refinement tools to capture backbone flexibility.

\section*{Methods accounting backbone flexibility}

\subsection*{\textbf{Molecular dynamics}}

\begin{figure}
\renewcommand{\familydefault}{\sfdefault}\normalfont
\centering
\includegraphics[scale=0.5]{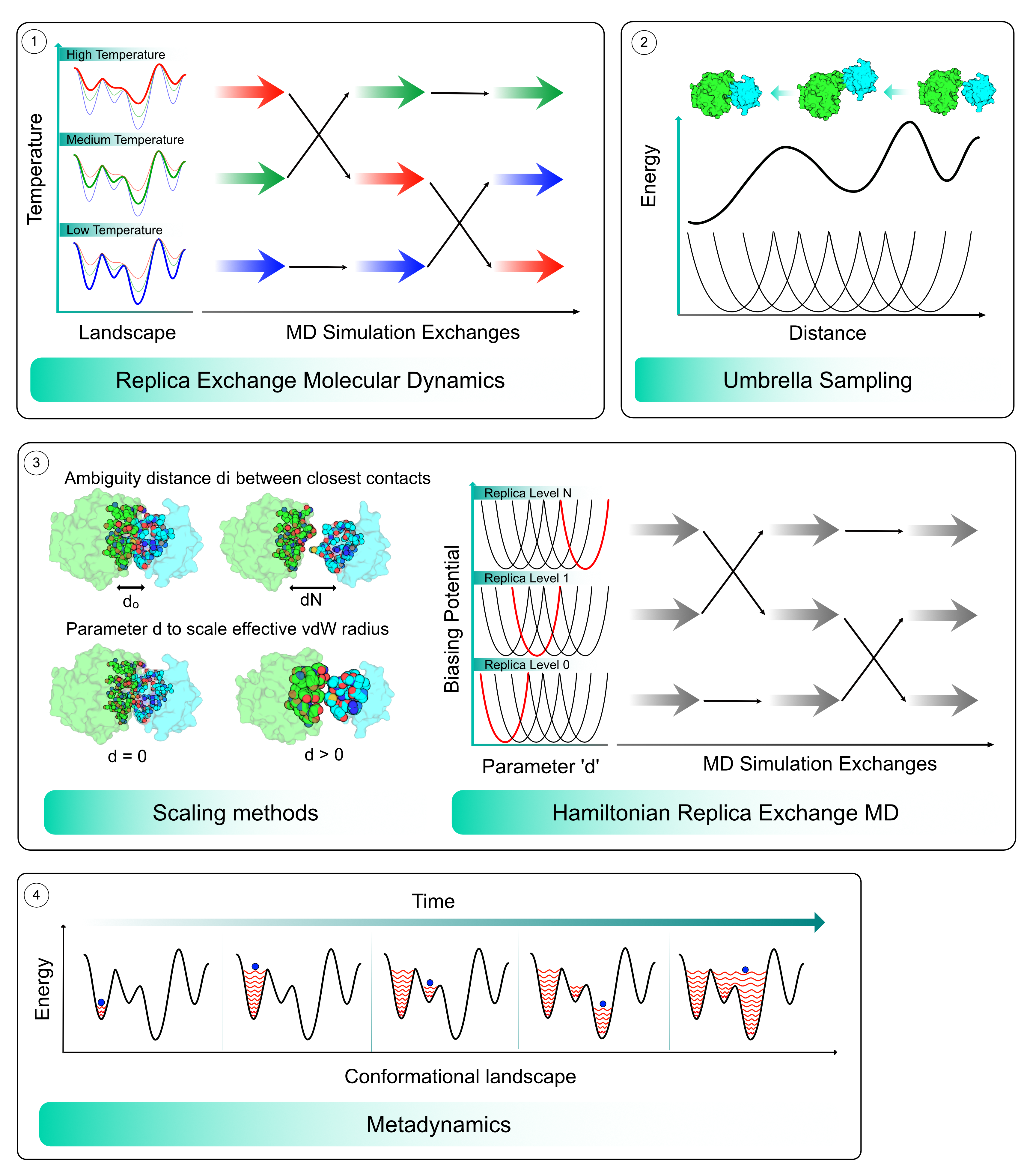}
\caption{\textbf{Enhanced Sampling approaches in protein docking.} 1. \textit{\textbf{Temperature replica exchange}} MD/MC approaches utilize temperature as the variable parameter across replicas \citep{Liu2005,Zhang2013}. The smoothening of the relatively rugged energy landscape enables sampling of distinct energy basins. 2. \textit{\textbf{Umbrella Sampling}} methods \citep{Kastner2011} split the reaction coordinate between an unbound and bound state into multiple windows. This enables biasing molecular dynamics trajectories along the reaction coordinate driving the system from one thermodynamic state to another. 3. \textit{\textbf{Hamiltonian replica change}} approaches introduce a biasing potential which can be either time-dependent, contact-dependent \citep{Ostermeir2017}* or geometry dependent \citep{Siebenmorgen2020}**. \textbf{Scaling methods} Top: Use of contact-dependent ambiguity constraints between protein partners. The weighted distance of the closest contacts of the partners defines bias potentials; Bottom: Bias based on increase in the effective pairwise vdW radii (an illustration to indicate the variable vdW radii across replicas for hamiltonian-based tempering); \textbf{Hamiltonian REMD:} The exchange trajectories with the biasing harmonic potential(red) and the range of potentials used across all the replicas in the system. 4. \textit{\textbf{Conformational flooding / Metadynamics}} utilize an exhaustive search within a local scope by introducing a funnel-shaped constraint potential \citep{Limongelli2013}. Short metadynamics simulations have been equipped to obtain backbone conformations for ensemble-docking \citep{Basciu2019}.}
\label{fig:EnhancedSampling}
\vspace{-19.42863pt}
\end{figure}

Molecular dynamics (MD) is one strategy that is often used after grid-search or template-based approaches for refinement (Figure \ref{fig:EnhancedSampling}) \citep{Dapkunas2018,Christoffer2019}. 
Unbiased, all-atom MD simulations can provide a high-resolution, time-resolved microscopic model of protein-protein interactions. MD calculates Newtonian trajectories using physics-based energy functions to simulate protein association and dissociation events. MD use for protein docking has been limited because non-native local minima trap proteins, and dissociation is too slow \citep{Shaw2010}. Over the past decade, two new modifications to capture conformational changes are steered molecular dynamics (SMD) \citep{Krol2007}, which utilizes external force constraints, and Markov sampling, which breaks a long MD simulation into multiple short trajectories \citep{Plattner2017}. To accelerate dissociation of protein partners at sub-optimal binding regions, Ostermeir \textit{et al.} developed a Hamiltonian replica exchange MD protocol (H-REMD) for protein docking \citep{Ostermeir2017}*. In H-REMD, biasing potentials are based on the shortest distance between protein partner atoms (defined as ``ambiguity restraints''). As the biasing potential and associated ambiguity restraints vary across replicas, associated protein partners in one replica are forced to dissociate in another. Pan \textit{et al.} simulated long timescales in a global search space for a benchmark set of five targets on the special purpose machine Anton \citep{Shaw2014, Pan2019}. Their ``tempered binding'' protocol updates energy function parameters throughout the simulation: a soft-core van der Waals intermolecular potential is scaled so that long-lived states are dissociated more frequently, improving the sampling efficiency \citep{Pan2019}**. Further, Pan \textit{et al.} found that proteins often follow a repeated dissociation and association pattern rather than probing continually along the surface for the native binding site. Siebenmorgen \textit{et al.} similarly scaled atomic repulsions with the vdW radii \citep{Siebenmorgen2020}**. They varied the vdW attraction energy across replicas relative to the Lennard-Jones and electrostatic interactions (owing to increased ligand-receptor atom distance). Compared to conventional MD methods, their simulations sampled native-like states 30\% more often; resulting in blind docking predictions within 5 \r{A} of native for moderately flexible targets. MD-based docking on proteins that move more than 2.2 \r{A} RMSD upon binding has not yet been reported.

\subsection*{\textbf{Monte Carlo methods}}

In contrast to MD approaches that target flexibility with Newtonian dynamics; Monte Carlo (MC) methods sample by random moves often followed by minimization (MCM) \citep{Gray2006,Vajda2013}. MC allows a wide variety of conformational move types to sample diverse conformations. MC algorithms have emulated the kinetic binding models, namely key-lock, conformer selection (CS) and induced-fit (IF) mechanisms \citep{Wang2007,Chaudhury2008,Zhang2017}. The CS model chooses protein backbones from a pre-generated ensemble, thus this approach has the advantage of docking one partner's conformations at a time. However, CS docking can fail if the ensemble is devoid of native-like backbone conformations \citep{Kuroda2016}. For targets with \rmsdBU\ up to 2.5 \r{A}, Zhang \textit{et al.} generated ensembles of 40 structures for MC-based docking \citep{Zhang2017}. This ensemble docking approach incorporates the ATTRACT coarse-grained protein model (Figure \ref{fig:cg_fft_ml}) \citep{Zacharias2005} in conjunction with replica-exchange (RE) to sample in backbone as well as rigid body space. Although the ensemble does not always include bound-like conformations of the proteins, their REMC-ensemble docking method obtains higher quality structures than MCM and REMC approaches. RosettaDock4.0 \citep{Marze2018}**, a conformer selection based MCM approach, modulates backbone swaps with an strategy that modulates rates of sampling of each conformer to handle ensembles of 100 structures for each protein partner (RosettaDock3.0 \citep{Chaudhury2008} docked from an ensemble of 10 structures). To diversify backbone conformations, the protocol generates monomer structures by three methods: (1) normal modes \citep{Atilgan2001} (2) backrub motions \citep{Smith2008} and (3) all-atom backbone refinement \citep{Tyka2011}. Further, to discriminate between near-native and non-native structures, they developed a more accurate coarse-grained energy function with 6-dimensional residue-pair data obtained from protein-protein interfaces in the Protein Data Bank (Figure \ref{fig:cg_fft_ml}) \citep{Fallas2017}. Marze \textit{et al.} report success on 49\% of moderately flexible and 31\% of flexible targets, the highest local-docking success rates yet reported \citep{Marze2018}**. 

\subsection*{\textbf{Sampling backbone conformations with normal modes}}

Since intrinsic fluctuations in proteins contribute to conformational change, some docking approaches utilize harmonic dynamics to capture protein backbone motions \citep{Zacharias1999,Grunberg2006,Zacharias2010}. Normal modes of vibration represent internal motions of a protein based on a Hookean potential between close residues. Normal mode analysis (NMA) is incorporated in docking approaches such as ATTRACT \citep{DeVries2013}, FiberDock \citep{Mashiach2010}, SwarmDock \citep{Moal2010} and EigenHex \citep{Venkatraman2012}. To mimic induced-fit, Schindler \textit{et al.} developed iATTRACT \citep{Schindler2015} by moving interface residues in Cartesian coordinate space subject to NMA-generated harmonic potentials. iATTRACT served as a refinement stage and improved the fraction of native contacts predicted by 70\%. For targets with unbound to bound interface RMSD over 4 \r{A}, iATTRACT can achieve acceptable quality models \citep{Schindler2015}. Population-based methods such as particle swarm optimization (PSO) have also employed NMA. PSO is a heuristic approach that  optimizes the multiple degrees of freedom using a set of multiple systems. The SwarmDock algorithm recently incorporated dynamic cross-docking \citep{Torchala2020}* of multiple backbone conformations within its PSO routine. It obtains an ensemble of conformational states of individual protein partners by using elastic network normal mode calculations and samples with the five lowest frequency non-trivial modes. SwarmDock achieved medium or high quality structures even for difficult targets with i-RMSD between 2.2 and 6 \r{A} along with a challenging prior CAPRI target (T136) \citep{Torchala2020, Lensink2019a}. Extending the swarm intelligence methods, the LightDock algorithm uses a ``glowworm'' swarm optimization to sample different backbone conformations in local regions of the protein surface with an anisotropic network model \citep{Jimenez-Garcia2018}. LightDock additionally uses multiscale modeling to combine all-atom and coarse grained scoring functions.

\begin{figure}[tb]
\renewcommand{\familydefault}{\sfdefault}\normalfont
\centering
\includegraphics[width=12cm]{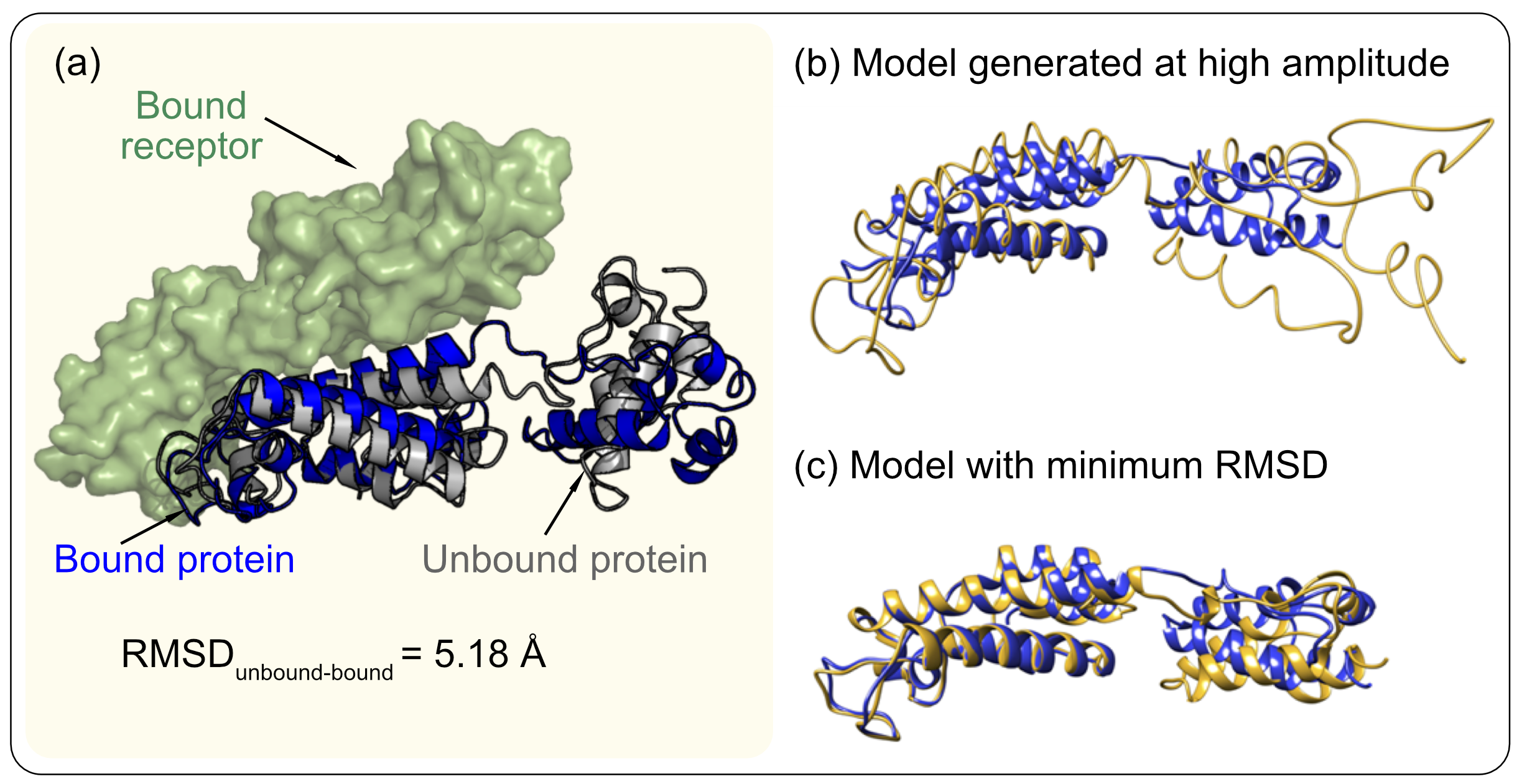}
\caption{\textbf{Internal coordinates NMA captures larger conformational change.} (a) Schematic of the bound homodimer (PDB ID: 2EIA) and unbound monomer (1EIA) forms of equine infectious anemia virus (EIAV) capsid protein p26 (\rmsdBU\ of the binding domain is 5.2 \Angstrom). (b) Model generated by internal coordinate NMA at maximum amplitude (yellow) retains realistic bond lengths and angles. (c) iNMA with the optimal mode magnitudes yields a structure within  3 \r{A} RMSD of the bound form. Panels (b) and (c) adapted from Frezza and Lavery (2019) \citep{Frezza2019}**. }
\label{fig:iNMA}
\end{figure}

While normal modes have typically been used on individual protein partners prior to docking, Oliwa and Shen introduced the complex NMA in docking to also sample molecular complex fluctuations \citep{Oliwa2015}. By calculating modes of an encounter complex, this approach focuses on the binding region as it reduces the dimensionality of the search space \citep{Chen2017}. One of the problems of NMA is that higher frequency modes often distort protein bonds. To overcome this limitation, Frezza and Lavery developed the internal coordinate NMA (iNMA) approach to move in the torsion angle space, that is, with fixed bond lengths and angles (Figure \ref{fig:iNMA}) \citep{Frezza2015}. With a reduced protein model in an internal coordinate space, they captured larger conformational changes from eigenvectors of low-frequency modes \citep{Frezza2019}**. iNMA can generate structures within 3 \r{A} of the bound state when starting from the unbound for 39\% of single-domain and 45\% of multi-domain proteins in their benchmark.

\subsection*{\textbf{Machine learning methods}}

Although protein folding has been one prime focus of deep learning methods in biology (e.g., AlphaFold \citep{Senior2019, Senior2020} and RaptorX \citep{Wang2017}), in recent years, a few studies have explicitly addressed challenges relevant to protein docking \citep{Gao2020}. Protein binding sites can be thought of as an information-rich molecular space that can be mined for elucidating protein interactions \citep{Shulman-Peleg2004,Fout2017,Zeng2018}.

One approach is to use this information to create score functions for use with traditional docking approaches. For example, Geng \textit{et al.}  used graph representations to train a support vector machine (SVM) on native and non-native protein complex structures to develop a scoring potential (GraphRank) to rank docked poses \citep{Geng2019}. And iScore, composed of the GraphRank and HADDOCK \citep{Dominguez2003} scores, achieved top performance in CAPRI scoring rounds (medium or high quality structures for nine out of 13 targets). 

Other teams have used deep learning techniques to identify protein interfaces by extrapolating image recognition tools to protein structures. RaptorX-ComplexContact \citep{Zeng2018} uses a deep residual neural network trained on single-chain proteins to predict contacts between binding partners, achieving the top contact prediction scores in CASP \citep{Kryshtafovych2019}. Another approach is to characterize interaction environments. Townshend \textit{et al.} created ``voxels,'' i.e., volumetric pixels with local atomic information for every protein surface residue, and with this 3D representation, they trained a deep 3D convolutional neural network (SASNet) on a curated database of bound protein complex structures \citep{Townshend2019}. Pittala \textit{et al.} employed graph convolutions with the nodes representing the amino acid residues and edges connecting residues with a $C_{\beta} - C_{\beta}$ distance under 10 \r{A} \citep{Pittala2020}. They placed geometric and chemical features on both nodes and edges and used a graph neural network to predict epitopes and paratopes in antigen-antibody interfaces. In a unique approach by Gainza \textit{et al.}, a geometric deep learning model (MaSIF) used molecular interaction ``fingerprints'' calculated using geometric and chemical features of protein surfaces \citep{Gainza2020}** (\ref{fig:cg_fft_ml}). Their deep network was composed from geodesic convolutional layers, and they used it to predict binding sites, evaluate alternate docked interfaces, and assess likelihood of a given protein-protein interaction. Relative to conventional rigid docking methods on protein targets, MaSIF-search can perform ultra-fast scanning to identify true `binder' with similar accuracy but significantly faster (4 CPU-minutes vs.\ 45 hours for PatchDock and 93 days for ZDOCK to evaluate a benchmark of 100 bound protein complexes). 

In a study to explore how neural networks might be used to generate structures with considerable backbone motion, Degiacomi trained an autoencoder with conformations from MD simulations, compressing the protein motion into a low-dimensional latent space \citep{Degiacomi2019}*. By training with simulations of both closed (bound) and apo conformations of a target protein, the autoencoder generated an intermediate closed-apo conformation at 0.8 \r{A} RMSD \citep{Degiacomi2019} from the native state. However, when the autoencoder was trained only with open conformations, the generator could only create structures far from the closed state (over 4.2 \r{A}), limiting the utility of this approach for blind docking. In an approach suitable for blind cases, Cao and Shen developed a Bayesian active learning (BAL) model to quantify uncertainty in protein structure quality, and then they extended their model to flexible protein docking \citep{Cao2020}*. The Bayesian framework determines the posterior probability as it samples backbone conformations \citep{Oliwa2015}. 
Flexibility is captured with low-frequency complex-NMA modes, and in principle it can be extended to higher frequencies that capture loop and hinge motions. Compared to ZDOCK \citep{Chen2003} and PSO, BAL improves the interface RMSD of the near-native predictions by 0.5 \r{A}. 

\section*{\textbf{Conclusions}}

In conjunction with experimental data, docking has advanced a range of biological and health applications (e.g., Alzheimer's disease \citep{Frost2020}, celiac disease \citep{Hoydahl2019}, SARS-CoV-2 \citep{Cleri2020}, influenza \citep{Xu2020}, cancer \citep{Kalin2018}, and heart disease \citep{Alford2020}, to name just a few). Over the past few years, docking success rates have improved on ``difficult'' blind prediction targets, but rates need to be higher for docking to be a reliable stand-alone tool in all cases. Clearly, a diverse and impressive array of tools has steadily advanced toward reliably capturing large conformational changes in protein docking. Docking will be even more impactful when the field finally overcomes this challenge.

\section*{Acknowledgements}

This work was supported by the National Institutes of Health through grant R01-GM078221. We thank Marc Lensink for generously providing us with data from CAPRI and Sai Pooja Mahajan and Sudhanshu Shanker for helpful comments on the manuscript.

\section*{Conflict of Interest}

Dr. Jeffrey J. Gray is an unpaid board member of the Rosetta Commons. Under institutional participation agreements between the University of Washington, acting on behalf of the Rosetta Commons, Johns Hopkins University may be entitled to a portion of revenue received on licensing Rosetta software including applications mentioned in this review. As a member of the Scientific Advisory Board, Dr. Gray has a financial interest in Cyrus Biotechnology. Cyrus Biotechnology distributes the Rosetta software, which may include methods mentioned in this review.

\bibliography{tackling-bb-flexibility}

\end{document}